\documentclass[AMA,STIX1COL]{WileyNJD-v2}
\usepackage[utf8]{inputenc}
\usepackage{amsfonts,epsfig}
\usepackage{hyperref}
\usepackage{breakurl}
\usepackage{tabu} 
\usepackage{booktabs}
\usepackage{outline}
\usepackage{algorithm}
\usepackage{multicol}
\usepackage{float}
 \usepackage{adjustbox}
 \usepackage{rotating}
 \usepackage{caption}
 \usepackage{subcaption}
\usepackage{graphicx,amsmath,multicol,multirow,amssymb,amsbsy,pifont}
\mathchardef\mhyphen="2D

\newcommand \lpar{\left(}
\newcommand \rpar{\right)}
\newcommand \lbr{\left[}
\newcommand \rbr{\right]}

\articletype{Research Article}

\received{26 April 2016}
\revised{6 June 2016}
\accepted{6 June 2016}

\begin{document}

\title{Embedded multilevel regression and poststratification: Model-based inference with incomplete auxiliary information\protect}

\author[1]{Katherine Li}

\author[2]{Yajuan Si*}

\authormark{Li and Si}

\address[1]{\orgdiv{Department of Biostatistics}, \orgname{School of Public Health, University of Michigan, Ann Arbor}, \orgaddress{\state{Michigan}, \country{USA}}}

\address[2]{\orgdiv{Survey Research Center}, \orgname{Institute for Social Research, University of Michigan, Ann Arbor}, \orgaddress{\state{Michigan}, \country{USA}}}

\corres{*Yajuan Si, ISR 4014, 426 Thompson St, Ann Arbor, MI 48105, USA. \email{yajuan@umich.edu}}

\abstract[Summary]{

Health disparity research often evaluates health outcomes across demographic subgroups. Multilevel regression and poststratification (MRP) is a popular approach for small subgroup estimation due to its ability to stabilize estimates by fitting multilevel models and to adjust for selection bias by poststratifying on auxiliary variables, which are population characteristics predictive of the analytic outcome. However, the granularity and quality of the estimates produced by MRP are limited by the availability of the auxiliary variables’ joint distribution; data analysts often only have access to the marginal distributions. To overcome this limitation, we embed the estimation of population cell counts needed for poststratification into the MRP workflow: embedded MRP (EMRP). Under EMRP, we generate synthetic populations of the auxiliary variables before implementing MRP. All sources of estimation uncertainty are propagated with a fully Bayesian framework. Through simulation studies, we compare different methods and demonstrate EMRP's improvements over alternatives on the bias-variance tradeoff to yield valid subpopulation inferences of interest. As an illustration, we apply EMRP to the Longitudinal Survey of Wellbeing and estimate food insecurity prevalence among vulnerable groups in New York City. We find that all EMRP estimators can correct for the bias in classical MRP while maintaining lower standard errors and narrower confidence intervals than directly imputing with the WFPBB and design-based estimates. Performances from the EMRP estimators do not differ substantially from each other, though we would generally recommend the WFPBB-MRP for its consistently high coverage rates.
}

\jnlcitation{\cname{%
\author{K. Li} and
\author{Y. Si}} (\cyear{2022}), 
\ctitle{Embedded Multilevel Regression and Poststratification: Model-based Inference with Incomplete Auxiliary Information}, \cjournal{Statistics in Medicine}, \cvol{2022;00:1--6}.}

\maketitle

\footnotetext{\textbf{Abbreviations:} EMRP, embedded multilevel regression and poststratification; LSW, Longitudinal Survey of Wellbeing; MRP, multilevel regression and poststratification; WFPBB, weighted finite population Bayesian bootstrap}

\section{introduction}\label{sec1}

Health disparity research often evaluates health outcomes across demographic subgroups, with a focus on vulnerable minors.\citep{cdc:jacobs} Identification of such at-risk groups provides essential information for health policy research. Multilevel regression and poststratification (MRP) is a method that has recently become popular for subgroup estimation, as it can extrapolate sample inferences to the target population with either probability or nonprobability samples.\cite{BNFP:SI15,prior-si2018,mrp-si20,mrp-covid21, mrp-covid22} MRP has two key components: 1) small subgroup estimation by fitting a predictive multilevel outcome model with a large number of covariates and regularizing with Bayesian prior specifications; and 2) poststratification to adjust for selection bias. The flexible modeling of analytic outcomes can capture complex data structures conditional on poststratification cells---which are determined by the cross-tabulation of categorical variables that affect the sample inclusion (selection and response)---and the use of population control information in poststratification can balance the sample discrepancy.\citep{hs79,gelman:carlin:00} 

The availability of population control information that is strongly related to the analytic variables affects the inferential validity for both model-based and design-based approaches. Poststratification is crucial for sampling selection and nonresponse bias adjustment and requires the use of predictive auxiliary variables with their joint distribution in the population. Practical applications solicit population information from either census records or large studies with minimal errors. For example, across MRP application studies, Si et al.\cite{prior-si2018} obtain the joint population control distribution from the American Community Survey (ACS); Wang et al.\cite{wang:gelman14} use the aggregated exit polls; Zhang et al.\cite{Zhang15-mrp} use census records; Yougov\citep{Yougov} uses the Current Population Survey; and Ghitza and Gelman\cite{votedatabse:mrp2020} turn to large-scale voter registration databases to directly obtain such information for the poststratification adjustment. 

However, the population joint distribution of poststratification variables (i.e., poststratifiers) is often unavailable, resulting in unknown cell counts; we may only have the marginal distributions of partial variables. We formalize an extension of MRP by embedding the estimation of population cell counts with incomplete auxiliary information, a framework we refer to as ``embedded MRP'' (EMRP). These estimated cell frequencies are derived from synthetic populations generated from nonparametric bootstrap and sequential imputation approaches, and all sources of estimation uncertainty are propagated under a Bayesian paradigm. EMRP is a class of model-based strategies for accounting for survey weights in modeling as well as a data integration framework for combining multiple sources of data.

Our primary contributions are that we consolidate different methods under the EMRP class of estimators, implement them to demonstrate the EMRP integrative workflow, and compare their performances with alternatives. We consider three methods for estimating the population cell counts: 1) the weighted finite population Bayesian bootstrap (WFPBB),\cite{fpbb14} 2) draws from a multinomial distribution informed by observed sampled cell frequencies,\cite{BNFP:SI15,Makela-sm18} and 3) predictive values from a logistic or multinomial-logit regression that explicitly models the distribution of the missing poststratifier on other, fully observed variables.\cite{reilly:gelman:katz01,unknownx:mrp:Phillips15} While multinomial draws and logistic regression predictions have established literature use in cell count estimation, this is the first instance (to our knowledge) that the WFPBB has been used for this purpose. 

Our methodological research is motivated by the practical operation of an ongoing survey: the New York City (NYC) Longitudinal Survey of Wellbeing (LSW),\cite{RHreport} which aims to provide assessments of poverty, material hardship, and child and family general health and wellbeing of the NYC residents. The survey organizers are particularly interested in the life quality aspects of minority groups. The survey collects data from NYC adult residents by oversampling from low-income neighborhoods and following up every three months. We have calibrated the baseline samples to the 2011 ACS-NYC records, assuming all variables affecting the sample inclusion are available in the ACS data: sex, age, race, education, and income.\citep{RHweighting}

We are interested in applying MRP to estimate food insecurity prevalence for sociodemographic subgroups. Since food insecurity is associated with an increased risk of diabetes and hypertension in adults, it is of interest to healthcare professionals and policymakers to identify subpopulations at risk for food insecurity to improve health outcomes.\citep{doi:10.1377/hlthaff.2015.0645} Weighted design-based estimation can inflate estimation variability---especially for small groups---and we would likely obtain more stable estimates with MRP. Considering residents who often visit food acquisition agencies in NYC tend to suffer from food insecurity and material hardship, in addition to having a different sample inclusion rate, MRP inferences with the LSW sample should adjust for visiting frequency. However, we are unable to do so as the population distribution of visiting frequency is unknown.

The problems we face in the LSW baseline survey are reflective of problems in most survey applications. When combining multiple data sources, we often encounter incomplete auxiliary information. We would like to incorporate the variables that affect sample inclusion into the survey variable model by integrating the estimation of poststratifier distributions under EMRP. We consider nonparametric bootstrap algorithms and parametric models for the population cell count estimation.

The paper structure is organized as follows. Section~\ref{method} outlines the methods that fall within the EMRP framework. Section~\ref{simulation} provides a simulation study to evaluate the performances of different EMRP methods. We apply EMRP to estimate food insecurity prevalence using the LSW dataset in Section~\ref{application}. Finally, Section~\ref{discussion} summarizes the findings and extensions.

\section{Methods}
\label{method}

Suppose the outcome in the population is $Y_i$, and the auxiliary variables of the population are categorical (or discretized continuous variables) denoted by $X_i$ and $Z_i$, for $i=1,\dots, N$, where $N$ is the population size and either $X_i$ or $Z_i$ can be multivariate. For illustrative purposes, we assume that $X_i$ is univariate and that $Z_i$ represents multiple variables. The sample of size $n$ drawn from the population includes $(z_i, x_i, y_i)$, for $i=1, \dots, n$. 

When the population distribution of $Z$ is known, the cross-tabulation of $Z$ results in $M$ cells with known cell sizes $N^z_m$'s with $\sum^M_{m=1} N^z_m = N$.  If our goal is to estimate the overall population mean of the outcome $\theta =\sum_iY_i/N$, then the classical MRP estimator can be described as follows:

\begin{equation}
\hat{\theta}^z_{MRP}=\sum^M_{m=1} \frac{N^z_m}{N}\hat{\theta}^z_m,
\label{eq:mrp}
\end{equation}
where $\hat{\theta}^z_m$ is the model-based estimate in cell $m$ for $m=1,\dots, M$. The outcome model $(Y \mid Z)$ fitted to the sample data $(y_i,z_i)$ can be a Bayesian multilevel regression. Examples include a Bayesian hierarchical model with weakly informative or informative prior specifications, or a flexible prediction algorithm.\citep{prior-si2018,wang:gelman14,Ghitza:gelman-13,Carlin:AJE2018} With a binary outcome $y_i \in \{0,1\}$, the model can be a Bayesian logistic regression, Gaussian process regression model,\citep{BNFP:SI15} or stacked regression\citep{stackMRP20} by assuming that $y_i \sim \textrm{Bernoulli}(\theta_{j[i]})$, where $\theta_{j[i]} =\textrm{Pr}(y_i=1)=f(\alpha^z_i)$ is a function of parameters $\alpha^z_i$ corresponding to the level of $z$ for unit $i$ and cell $j[i]$ that unit $i$ belongs to.

However, the poststratifiers' information can be incomplete. The cross-tabulation of all the poststratifiers $(Z_i, X_i)$ results in $J$ cells. If the joint population distribution of $(Z_i, X_i)$ is unknown, then the population cell sizes must be estimated by $\hat{N}_j$'s, with $\sum_{j=1}^J\hat{N}_j=N$. We can embed the estimation of the poststratification weights into the classical MRP workflow and present the conceptual EMRP workflow in Figure~\ref{fig:emrp}. 
\begin{equation}
\label{emrp-m}
\hat{\theta}_{EMRP}=\sum^J_{j} \frac{\hat{N}_j}{N}\hat{\theta}_j.
\end{equation}

The differences between MRP~\eqref{eq:mrp} and EMRP~\eqref{emrp-m} estimators are as follows.

\begin{enumerate}
 \item The poststratification variables of EMRP are $(Z, X)$, while MRP poststratifies only on $Z$. Poststratifiers should be predictive of either the analytical outcome (primarily) or the response propensity (secondarily).\cite{mrp-si20,little-weighting-SM2005} We assume that the sample inclusion mechanism depends only on $Z$ such that the data are missing at random (MAR). If sample inclusion depends on incomplete auxiliary variables, then the data are missing not at random (MNAR) and both MRP and EMRP estimators are subject to bias. Nevertheless, EMRP is expected to reduce the bias of MRP by leveraging the correlation structure between $X$ and $Z$ to generate the joint distribution, which will be elaborated on below.
 
 \item EMRP estimates the population cell size $\hat{N}_j$ and propagates its uncertainty, while MRP treats $N^z_m$ as fixed. 
 The propagated variance estimate of EMRP is larger than that of the MRP estimator with known $N_j$, which is expressed in the following decomposition:

 \begin{align*}
\label{emrp-var}
var(\hat{\theta}_{EMRP}) & =var\lbr E\lpar \sum_{j} \frac{\hat{N}_j}{N}\hat{\theta}_j\mid  \hat{N}_j\rpar\rbr + E\lbr var\lpar \sum_{j} \frac{\hat{N}_j}{N}\hat{\theta}_j\mid  \hat{N}_j\rpar\rbr \\
& =var\lbr \sum_{j} \frac{\hat{N}_j}{N} E\lpar \hat{\theta}_j\mid  \hat{N}_j\rpar\rbr   + E\lbr var\lpar \sum_{j} \frac{\hat{N}_j}{N}\hat{\theta}_j\mid  \hat{N}_j\rpar\rbr,
\end{align*}
where the second term is the variance of the MRP estimator poststratifying on $(Z,X)$: $var(\hat{\theta}^{z,x}_{MRP})  = var\left(\sum_{j} \frac{N_j}{N}\hat{\theta}_j\right)$ with known $N_j$.

\item The cell estimate $\hat{\theta}_j$ under EMRP is based on the model $(Y \mid Z, X)$ fitted to the sample data $(y, z, x)$, while classical MRP fits the model $(Y \mid  Z)$ to the sample data $(y, z)$. Considering the subpopulation mean $E(Y \mid Z) = \sum_X E(Y \mid Z, X) \textrm{Pr}(X \mid Z)$, where $\textrm{Pr}(X \mid Z)$ is the conditional probability, the MRP estimator can be similar to the EMRP estimator asymptotically or when the sample resembles the population distribution of $X$ given $Z$ in the subgroup to achieve the equality based on the integration. However, because of small subgroup sizes in the sample data, it is possible that the observed frequency of $X$ values within a subgroup defined by $Z$ is different from the population distribution and the equality will not hold. We expect that the estimation of $E(Y \mid Z, X)$ will be more accurate than $E(Y \mid Z)$ when $X$ is a strong predictor of $Y$.  

\end{enumerate}

Hence, we expect that the EMRP estimator can reduce the bias of the MRP estimator, especially for small subgroups. We compare three strategies to estimate $\hat{N}_j$ based on $N^z_m$ and the sample data $(y, z, x)$: 1) drawing synthetic populations of $(Z, X)$ from the weighted finite population Bayesian bootstrap (WFPBB-MRP), 2) drawing $\hat{N}_j$ from a multinomial distribution (Multinomial MRP), and 3) predicting $X$ values in the population by regressing $X$ on $Z$ with logistic or multinomial-logit models (Two-stage MRP).  In all of our methods, the estimation uncertainty of $\hat{N}_j$ is propagated under a Bayesian framework.

\subsection{The weighted finite population Bayesian bootstrap (WFPBB-MRP)}

\begin{figure}[ht!]
    \centering
    \includegraphics[height=3in]{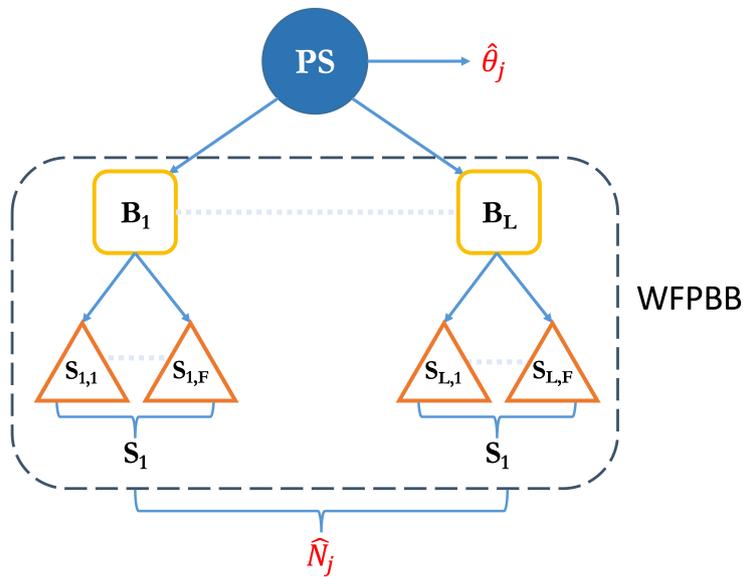}
    \caption{The illustration of the weighted finite population Bayesian bootstrap (WFPBB). From the original, or ``parent sample'' (PS), we generate Bayesian bootstrap samples ($B_1, \ldots, B_L$), and for each bootstrapped sample we pool $F$ populations drawn from the weighted P\'{o}lya to produce a single synthetic population ($S_l$) of size $F*N$.}
    \label{fig:wfpbb}
\end{figure}

Dong et al.\cite{fpbb14} propose the WFPBB as a nonparametric method of generating synthetic populations that can be analyzed as simple random samples by ``undoing'' the complex sampling design and accounting for the sampling weights. We use the WFPBB to estimate the joint distribution $(Z, X)$ in the population. Let $n^z_m$ denote the number of sampled units with $z=m$, and we have $\sum^M_{m=1} n^z_m=n$. We construct the sampling weights for unit $i$, $w_i=N^z_{m[i]}/n^z_{m[i]}$, where $m[i]$ represents the value of $z$ that is assigned to unit $i$, for $i=1,\dots, n$. The idea is to draw from the posterior predictive distribution of non-observed (nob) data given the observed (obs) data and weights: $(z_i,x_i)_{nob}\mid  (z_i,x_i)_{obs}, w_i$. Assume that there is a finite number of unique pairs $(z_i,x_i)$ in the sample, the population is also comprised of these unique pairs, and the corresponding counts for each pair in the population follow a multinomial distribution. Given a non-informative Dirichlet prior distribution on the multinomial probabilities, the P\'{o}lya distribution can be used in place of the Dirichlet-multinomial distribution to draw predictive samples and reduce the computational burden. We adapt and embed the WFPBB in the MRP implementation as ``WFPBB-MRP'' shown in Figure~\ref{fig:wfpbb}, following the steps below. 

\begin{enumerate}
    \item \textbf{Resample via Bayesian bootstrap (BB)}\cite{rubin:bb:81}:  To capture the sampling variability of drawing from the posterior distribution of the population parameters given the data from the ``parent'' (original) sample, we generate \textit{L} number of BB samples: $B_1,\ldots, B_L$, each of size $n$.
    \item \textbf{Recalibrate weights}: For each BB sample $B_l$, we recalibrate the bootstrap weights by multiplying the base weights by the number of replicates for unit $i$ in $B_l$ ($r^{l}_{i}$) and normalizing the weights to sum to the population size $N$, so that $w^l_i=N\frac{w_ir^{l}_{i}}{\sum_j w_jr^l_j}$.
    \item \textbf{Use the WFPBB to incorporate weights}:  Construct the initial P\'{o}lya urn based on $(z_i,x_i)$ with their corresponding replicate weights $w^l_i$'s and draw $N-n$ units with probability
    \begin{equation}
       \frac{w^l_i - 1 + l_{i, k-1}(N-n)/n}{N-n + (k-1)(N-n)/n},
        \label{eqpolya}
     \end{equation}
for the $k$th draw, $k \in \{1, \ldots, (N-n)\}$, where $l_{i, k-1}$ is the number of bootstrap selections of $(z_i,x_i)$ among the elements present in our urn at the $k-1$ draw. The draws form the WFPBB sample $S_{l,f}$ of size $N$. We repeat this step $F$ times, yielding multiple samples $S_{l,1},\ldots,S_{l,F}$---which we pool to create the synthetic population $S_l$--- of size $F*N$, where $S_l$ is considered a single draw from the WFPBB.
  
\item \textbf{Make inference}: For each of the samples $S_{l,f} \in \{S_{l,1},\ldots,S_{l,F}\}$ we obtain estimates $\hat{N}_j^{(l,f)}$. The estimate of $\hat{N}_j$ associated with synthetic population $S_l$ is the average of the estimates from the $F$ samples. We normalize the sum of $\hat{N}_j$ to be $N$ and obtain $L$ samples of $\hat{N}_j$. As a parallel step, we obtain posterior samples of $\hat{\theta}_j$ from the multilevel regression. The embedding of the two estimates yields the posterior samples of $\hat{\theta}_{EMRP}$; see Figure~\ref{fig:emrp}.
\end{enumerate}

In the WFPBB implementation, we have constructed weights based on the population distribution of $Z$. In practice, if the sampling weights are available from complex survey data, the weighted counts of the joint cells of $(Z, X)$ can be used to estimate their population totals, but the point estimate has ignored the sampling uncertainty of the survey. Similar to replication approaches that are often recommended to propagate the sampling variance, WFPBB undoes the weights in the generation of multiple synthetic populations and obtains the posterior samples of the population counts.

An alternative method for the population mean estimation is directly imputing the outcome of $Y$ with the WFPBB in a manner similar to that of Dong et al.\cite{fpbb14} The motivation of our adaptation is to improve estimation accuracy and precision using predictive auxiliary information. Direct WFPBB-imputation inference tends to become unstable with large standard errors when the sampled cell count is small or the replicate weights are highly variable; in such cases, the synthetic populations are mainly informed by a few outcome values observed in the sample and require a large number of replicates to be able to represent the true population. By using predictive auxiliary information in a multilevel model, we borrow inference across cells to improve estimation accuracy and precision. We compare the performance of the WFPBB-MRP to that of the direct WFPBB implementation in the simulation studies. 

\subsection{Multinomial draws (Multinomial MRP)}

Using the WFPBB can be computationally demanding. An alternative is to approximate the weighted Bayesian bootstrap by drawing the cell counts from a probability distribution---for example, a Poisson or multinomial distribution---as in Si et al.\cite{BNFP:SI15} and Makela et al.\cite{Makela-sm18} Suppose the possible values of the categorical variable $X$ (which can be multivariate) are $(1, \dots, C)$, where $C$ is the total number of levels for $X$. Here, we consider a multinomial distribution for $X$ conditional on the fully observed $Z$.

\begin{align}
\label{mnomial}
\nonumber X=(1, \dots, C)\mid Z=m &\sim \textrm{Multinomial}(N^z_m\mbox{; }  p_1,\dots, p_{C})\mbox{,  }\\
 p_c&=n^{z,x}_{m,c}/n^z_m\mbox{, } c=1,\dots,C.
\end{align}
We use the marginal population counts $N^z_m$ and the observed frequency in the sample $n^{z,x}_{m,c}/n^z_m$, where $n^{z,x}_{m,c}$ denotes the sample count of units with $(z_i=m, x_i=c)$, for $m=1,\dots, M$ and $c=1,\dots, C$. This distribution accounts for the correlation among variables based on the sample data. We extend MRP with synthetic poststratification (Multinomial MRP) to address the estimation of the joint distribution of $(Z, X)$.

Leemann and Wasserfallen\cite{mrsp:leemann17} assume both population margins of $(Z, X)$ are available (whereas any population information for $X$ is missing in our setting) and fit a multinomial model similar to~\eqref{mnomial}, except that the observed frequencies $n^{z,x}_{m,c}/n^z_m$ are updated to match the known margins of $X$. The approach taken by Leemann and Wasserfallen\citep{mrsp:leemann17} is similar to raking when population margins of multiple variables are available,\citep{deming40,rake:little91,BayesRake18} but draws repeated bootstrap samples to account for uncertainty in the synthetic poststratification. Multinomial MRP is a fully-Bayesian procedure that uses posterior samples to propagate all sources of estimation uncertainty.

\subsection{Sequential regression models (Two-stage MRP)}

Reilly et al.\cite{reilly:gelman:katz01} apply regression models to predict the unknown population poststratifier. Since MRP essentially predicts outcomes in the population, Kastellec et al.\cite{unknownx:mrp:Phillips15} propose a sequential estimation procedure by using two separate MRP procedures, i.e., Two-stage MRP, where the first stage estimates the population cell counts $\hat{N}_{j}$ using a multilevel model for $(X \mid Z)$, and the second estimates the cell means $\hat{\theta}_{j}$ with $(Y \mid Z, X)$. The model $(X \mid Z)$ assumes the probability $p_c$ in Model~\eqref{mnomial} is a function of pre-specified main effects or high-order interaction terms of $Z$ variables, but not necessarily the cell-wise observed frequencies across the cross-tabulation, resulting in regularized estimates with a strong dependency on model specification. If $X$ is a binary indicator (with values $0/1$), the first stage regresses $X$ on $Z$.
\begin{align}
\label{eq:2stage-xz}
 \textrm{logit(Pr}(X_i=1\mid Z_i, \beta_i^z)) &= \beta_i^z,
\end{align}
where the coefficient $\beta_i^z$ denotes the main or high-order effect corresponding to the level of $z$ for unit $i$. The synthetic predictions $(X_i\mid Z_i)$ would yield an estimate of the joint frequency in cell $j$, i.e., $\hat{N}_j$, for $j=1,\dots, J$. This method becomes cumbersome when $X$ has more than two levels, demanding a multinomial regression model.

The three EMRP strategies presented rely on the availability of population cell frequencies of $Z$ variables to generate synthetic populations of $(Z,X)$ and assume that the relationships between $X$ and $Z$ are the same between the sample and the population. The WFPBB-MRP constructs its base weights from the cross-tabulation of $Z$ variables and draws nonparametric Bayesian bootstrap samples. The Multinomial MRP assumes that the conditional distribution $(X\mid Z)$ within each cell in the cross-tabulation of $Z$ variables is a multinomial distribution with probabilities set as the observed frequencies. The Two-stage MRP applies a Bayesian multilevel model to predict $X$ given $Z$, where the model in practice often includes only the main effects of $Z$. 

The Two-stage MRP is subject to model misspecification and has stronger modeling assumptions than the WFPBB-MRP and Multinomial MRP, which automatically consider high-order interaction terms between $X$ and $Z$. Both the WFPBB-MRP and Multinomial MRP use the observed conditional distributions and may exhibit robust performances with large sample sizes. However, if the sample cell counts in the contingency table of $Z$ are sparse the estimated $\hat{N}_j$ from these two methods may have large variances, though the parametric Multinomial MRP has intrinsic variances that may be smaller than those from the WFPBB-MRP.  It is possible that some $Z$ values are not present in the sample, resulting in empty cells. While WFPBB-MRP and Multinomial MRP must omit empty cells, Two-stage MRP can generate the predictions of empty cells in the population since it relies on the model structure rather than the observed frequencies.
All methods propagate the uncertainty estimating the unknown population counts $\hat{N}_j$ in the EMRP variance estimator $var(\hat{\theta}_{EMRP})$, but there is a slight difference. WFPBB-MRP accounts for both the sampling variability of the external data $Z$ and the modeling uncertainty of estimating $(X \mid Z)$. Multinomial-MRP and Two-stage MRP approaches only account for the uncertainty of estimating $(X \mid Z)$ while treating the external distribution of $Z$ as fixed. Previous work \cite{mrsp:leemann17,reilly:gelman:katz01,unknownx:mrp:Phillips15} also ignores the sampling error of $Z$ and only accounts for the modeling uncertainty of $(X \mid Z)$. Since the external data $Z$ are often of large sample size---as is the case in the ACS---the counts are approximated as the population data with negligible sampling error. However, the sampling error would become substantial if the external sample size is small.

\section{Simulation studies}
\label{simulation}

We conduct simulation studies to compare embedded MRP methods, direct WFPBB imputation of the outcome, and classical MRP using \eqref{eq:mrp} for the overall population and subdomain inferences, both between the classes of estimators (WFPBB vs. classical MRP vs. EMRP overall) and within the class of EMRP estimators (WFPBB-MRP vs. Multinomial MRP vs. Two-stage MRP).  The simulation code is publicly available \href{https://github.com/likat/EMRP}{here}.

 \subsection{Setup}
\label{sim-setup}
We simulate a population of size $N=10,000$ with a binary outcome $Y$ and four categorical variables for poststratification: three $Z$ variables with known population distributions and one binary $X$ variable (0/1) without population information. The first two $Z$ variables ($Z_a$ and $Z_b$) have five levels and are generated from multinomial distributions, and the third variable ($Z_c$) is binary and drawn from a binomial distribution, where probabilities are normalized random numbers drawn from the uniform distribution, $\textrm{Uniform}(0.25, 1)$. The cross-tabulation of $Z$ variables results in $M=50 (= 5 \times 5 \times 2)$ cells. The variable $X$ is generated conditional on $Z$ with probability
\begin{align}
\label{x-z}
\textrm{Pr}(X_i=1\mid  Z_{ia}, Z_{ib}, Z_{ic}) = \textrm{expit}(\beta_0 + \beta_i^{Z_{a}}+ \beta_i^{Z_{b}} + \beta_i^{Z_{c}} + \beta_i^{Z_{a},Z_{c}}+ \beta_i^{Z_{b},Z_{c}}),
\end{align}
where the function $\textrm{expit}(\mu) = \textrm{exp}(\mu)/(1+\textrm{exp}(\mu))$, $(Z_{ia}, Z_{ib}, Z_{ic})$ denote the values of $(Z_{a}, Z_{b}, Z_{c})$ for unit $i\in \{1,\ldots, N\}$, respectively, and $\beta_i^{var}$ represents the value of $\beta^{var}$ corresponding to the level of $var$ for unit $i$, for $var \in \{Z_{a}, Z_{b}, Z_{c}\}$. We assume that Model~\eqref{x-z} has both main effects and interaction terms (INT) with $\beta_0 = -0.5$, $\beta^{Z_a} = (1.7, 0.25, 0.2, -0.75, -1.7)^\top$,  $\beta^{Z_b} = (2.3, 1.5, 0.15, 0.2, 0.9)^\top$, 
$ \beta^{Z_c} = (0, -1)^\top$, $\beta^{Z_a,Z_c} = (0, -0.6, 0.5, 0.35, -0.4)^\top$, and $\beta^{Z_b,Z_c} = (0,1.7,0.1,2,-0.75)^\top$. We also consider the case when Model~\eqref{x-z} has only main effects of $Z$ and include the output in Appendix~\ref{appendix:main}.

We specify the data generating process (DGP) for $Y$ as:
\begin{align}
\label{y-xz}
(Y_i\mid   Z_{ia}, Z_{ib}, Z_{ic}, X_i)\sim \textrm{Bernoulli}(\textrm{expit}(\alpha_0 + \alpha^{Z_a}_{i} +  \alpha^{Z_b}_{i} + \alpha^{Z_c}_{i} +\alpha^{X}_{i})).
\end{align}
The values are assigned as $\alpha_0 = 0$, $\alpha^{Z_a}=(1.37, -0.56, 0.36, 0.63, 0.40)^\top$, $\alpha^{Z_b}=(-0.11, 1.51, -0.09, 2.02, -0.06)^\top$, $\alpha^{Z_c}=(0,0.24)^\top$, and $\alpha^x = (0,-1.3)^\top$, where $(\alpha^{Z_a}, \alpha^{Z_b})$ are drawn from a standard normal distribution.

We assume that the inclusion mechanism depends only on the fully observed $Z$. The inclusion probabilities $\textrm{Pr}(I=1\mid  Z)$ are based on the cross-tabulation of three $Z$ variables, $Z_{cat} \in \{1,\ldots, M\}$, with values drawn from different ranges. Table~\ref{inc-pr} gives the cell indices and ranges of inclusion probabilities. We randomly draw values with replacement from each interval of equally spaced probabilities and assign them to the corresponding cells. We draw 200 repeated samples from the population with the pre-specified inclusion mechanism $\textrm{Pr}(I=1\mid  Z)$. The population cell frequencies range from 3 to 470, with an average of 100. In the repeated studies, the resulting average overall sample size is 4104; we obtained similar results with a smaller average sample size of 2052.

We are interested in the overall population and subgroup mean estimates. We assume that the MAR sample inclusion mechanism depends only on $Z$. Within each poststratification cell, MRP assumes that the inclusion probability is the same, and the individuals are independently and identically distributed. EMRP uses the correlation between $X$ and $Z$ to infer the cell counts for poststratification with $(Z, X)$. The inclusion probabilities can therefore be treated as a summary statistic of the distribution of $Z$ and their correlation with the inclusion mechanism. We use the intervals of inclusion probabilities to create subgroups of interest. 

We consider two methods to create subgroups. First, we create four subgroups based on the percentiles of the inclusion probabilities of the $J$ cells and the distribution of $(X \mid Z)$. Each subgroup contains 20 cells: the first subgroup includes cells with inclusion probabilities lower than or equal to the 40th percentile; the second contains those with moderate inclusion probabilities between the 20th and 60th percentiles; the third group includes those with medium inclusion probabilities between the 40th and 80th percentiles, and the fourth group covers high inclusion probabilities above or equal to the 60th percentile. The $X$ categories have different frequencies across subgroups, where for the 1st, 2nd, and 4th subgroup, 15 of the 20 cells have $X=0$ and 5 have $X=1$; for the 3rd subgroup, 5 cells have $X=0$, and 15 have $X=1$.  The average sampled cell sizes in the $(Z, X)$ cross-tabulation table for subgroups with low $p_I$: 0.03-0.27, moderate $p_I$: 0.21-0.38, medium $p_I$: 0.30-0.54, and high $p_I$: 0.40-0.95 are (9, 24, 53, 87) with average subgroup sample sizes (198, 491, 1064, 1744). The simulation scenarios cover cases with sparse cells and small groups. 

Second, the four subgroups are based on cells in the cross-tabulation of only $Z$ variables that are fully observed poststratifiers. All parameters are identical to those in the first subgrouping definition except for the subsampling procedure in each inclusion probability bracket: subgroups are defined by a random sample of 10 cells based on the cross-tabulation of $Z$ variables in each bracket instead of creating an imbalance of $X$ categories. The average sampled cell sizes in the $Z$ cross-tabulation table for subgroups with low $p_I$: 0.03-0.25, moderate $p_I$: 0.25-0.38, medium $p_I$: 0.34-0.51, and high $p_I$: 0.40-0.93 are (13, 28, 40, 62) with average subgroup sample sizes (262, 567, 815, 1251).

Table~\ref{x-diff} gives the differences of the observed values in one random sample and the population values of the missing poststratifying variable's frequency distributions $\textrm{Pr}(X=1)$ within the subgroups of two cases: 1) subgroup membership is defined based on the joint $(Z,X)$ distribution and 2) membership is defined based on categories of $Z$ only, which shows that the first scenario generally has larger differences than the second scenario. We expect that larger differences in terms of $\textrm{Pr}(X=1)$ lead to more different EMRP and MRP estimates.

For all three EMRP methods, the outcome model fitted to the sample data is identical to the DGP of $Y$ (see \eqref{y-xz}). 

The outcome model for classical MRP omits the main effect for $X$. The estimation model for $(X\mid Z)$ in the Two-stage MRP only accounts for the main effects, as misspecified:
\[
\textrm{logit}(\textrm{Pr}(X_i=1\mid Z_i)) = \beta_0 + \beta_i^{Z_{a}}+ \beta_i^{Z_{b}} + \beta_i^{Z_{c}}.
\]

To implement the WFPBB, we use the \texttt{polyapost} package version 1.6~\citep{polyapost} to draw from the weighted P\'{o}lya posterior distributions and generate $L=1000$ synthetic populations of size $F * N = 20 * 10,000$ per sample for inferences. We use Stan for the fully Bayesian posterior computation with MRP and perform convergence diagnostics.~\citep{stan-software:2021} For each sample, we fit the outcome model using two Markov chain Monte Carlo (MCMC) chains with 2000 iterations and keep the last 500 iterations from each for a total of 1000 draws (permuted and merged across chains) for estimation. Regression coefficients of multiple categories are assigned weakly informative priors\cite{gelman2008weakly}: normal distributions with mean 0 and unknown standard deviation parameters that are assigned hyperpriors of $\textrm{Cauchy}^{+}(0, 1)$, where $\textrm{Cauchy}^+(0, 1)$ is the half-Cauchy distribution restricted to positive values with a standard deviation $1$. For the intercept and coefficients of binary predictors, we use noninformative priors.

We assess bias, root mean squared error (rMSE), the average length of 95\% confidence intervals (CI length), and the nominal coverage rate of 95\% CIs (coverage rates) for the finite population quantities of interest. To quantify the uncertainty surrounding the $\hat{N}_j$ estimation in EMRP, we pair each of the $1000$ sets of cell mean estimates $\hat{\theta}_j$ with one of $1000$ sets of $\hat{N}_j$. This means we construct the sets of $\hat{N}_j$ from $1000$ synthetic populations of size $F * N$ for WFPBB-MRP, $1000$ draws of $\hat{N}_j$ from the multinomial distribution for Multinomial MRP, and $1000$ posterior draws from the sampler fitting the logistic regression model for $(X\mid Z)$ for Two-stage MRP. We compute 95\% CIs of the estimates of all five methods for a given sample by taking the 2.5th and 97.5th percentiles of their respective $1000$ posterior estimates.

We also present the survey-weighted and unweighted mean estimators. We use the same base weights as those in WFPBB ($w_i=N^z_{m[i]}/n^z_{m[i]}$) and the \texttt{survey} package version 4.1-1~\cite{lumley2020a} to obtain the design-based estimates, standard errors with the finite population correction, and 95\% confidence intervals. \\

\subsection{Results}

\begin{figure}[ht!]
\centering
\includegraphics[width=0.9\linewidth]{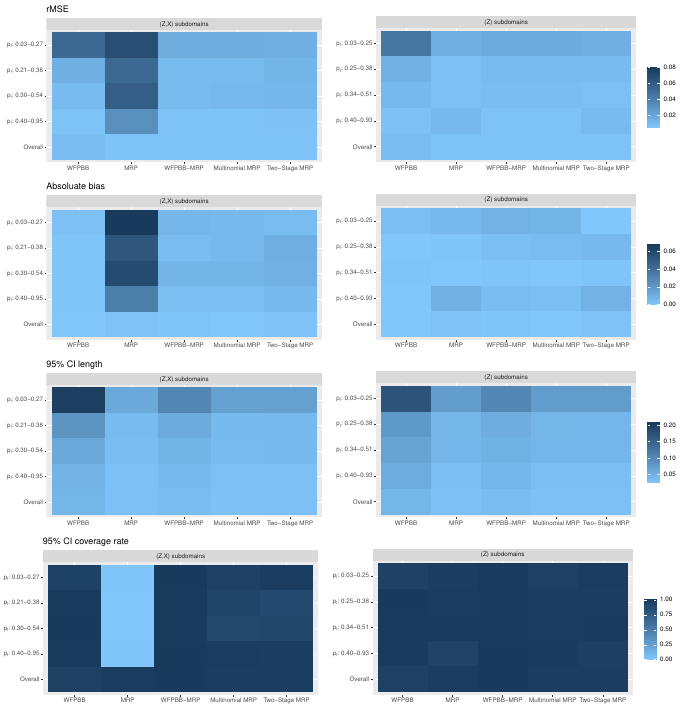}
\caption{Comparing the simulation cases for the overall and subdomain mean estimates, where subdomains are defined by inclusion probability ranges and either the joint distribution of $(Z,X)$ (left) or the levels of $(Z)$ (right). We compare the root mean squared error (rMSE), absolute bias, average 95\% confidence interval (CI) length, and 95\% CI coverage rates between the direct imputation of the outcome using the weighted finite population Bayesian bootstrap (WFPBB), classical multilevel regression and poststratification (MRP), and embedded multilevel regression and poststratification (EMRP) methods (WFPBB-MRP, Multinomial MRP, Two-stage MRP). Darker colors correspond to higher values.}
\label{fig:simresults_subsample}
\end{figure}

We present the simulation results by plotting heatmaps in Figure~\ref{fig:simresults_subsample} and reporting detailed values in Tables~\ref{tab:intxz} and~\ref{tab:intxz-nosub}, for the two subgrouping methods, respectively. 

When the subgroup is defined by the joint distribution of $(Z, X)$, as shown in the left column of Figure~\ref{fig:simresults_subsample}, the classical MRP yields high bias values and near-zero coverage rates due to the differences between the observed distribution of $(X \mid Z)$ and the population distribution in the subgroups. The EMRP estimators correct for these deficiencies: for subgroup estimates, classical MRP has absolute bias values of at least 0.033 and coverage rates of at most 0.005 while the EMRP methods produce absolute bias values of at most 0.011 and coverage rates of 0.870 or above. The EMRP methods do not differ substantially; the rMSE values differ by at most 0.003 and the bias values by 0.006. While WFPBB-MRP has a coverage rate of at least 0.99 in all subdomains, Multinomial MRP and Two-stage MRP have slight undercoverage in subgroups with moderate and medium inclusion probabilities $(p_I:0.21-0.38)$ and $(p_I: 0.30-0.54)$: (0.885, 0.895) and (0.910, 0.870), respectively.

Table~\ref{tab:intxz} shows that the survey-weighted estimator has competitive bias and coverage rates near or above 95\% except in the subgroup with low inclusion probabilities. Its rMSE and bias values are comparable to those from imputing the outcome with WFPBB, but its CIs are much narrower. The stabilizing benefit of Bayesian multilevel modeling is apparent when we compare direct imputation with WFPBB with EMRP methods. While the former yields comparable bias, the large variation between synthetic population inflates its rMSE and CI width compared to EMRP methods. This results in conservative coverage rates in most subgroups, though the bias in the low inclusion probability subgroup $p_I$: 0.03-0.27 incurred by the variable weights and sparse sampling results in slight undercoverage (0.940) despite wide intervals: the CI of the WFPBB estimate for the low group spans 0.202 compared to 0.105 from the WFPBB-MRP and 0.073 from the Multinomial-MRP and Two-stage MRP.

The right column of Figure~\ref{fig:simresults_subsample} shows that the MRP and EMRP estimates are similar for subgroups defined by the categories of $Z$ only, where the adjustment of incomplete $X$ in EMRP does not provide additional gains. Consistent with Table~\ref{x-diff}, the largest difference in the $\textrm{Pr}(X=1)$ is in the group with low inclusion probabilities ($p_I$: 0.03-0.25), for which the EMRP and MRP estimates have the most prominent dissimilarity among the five estimates. Table~\ref{tab:intxz-nosub} shows that WFPBB and survey-weighted estimators tend to have larger variances than the EMRP and MRP estimates, especially for the subgroup with low inclusion probabilities.

\begin{figure}[ht!]
\centering
\includegraphics[width=0.75\linewidth]{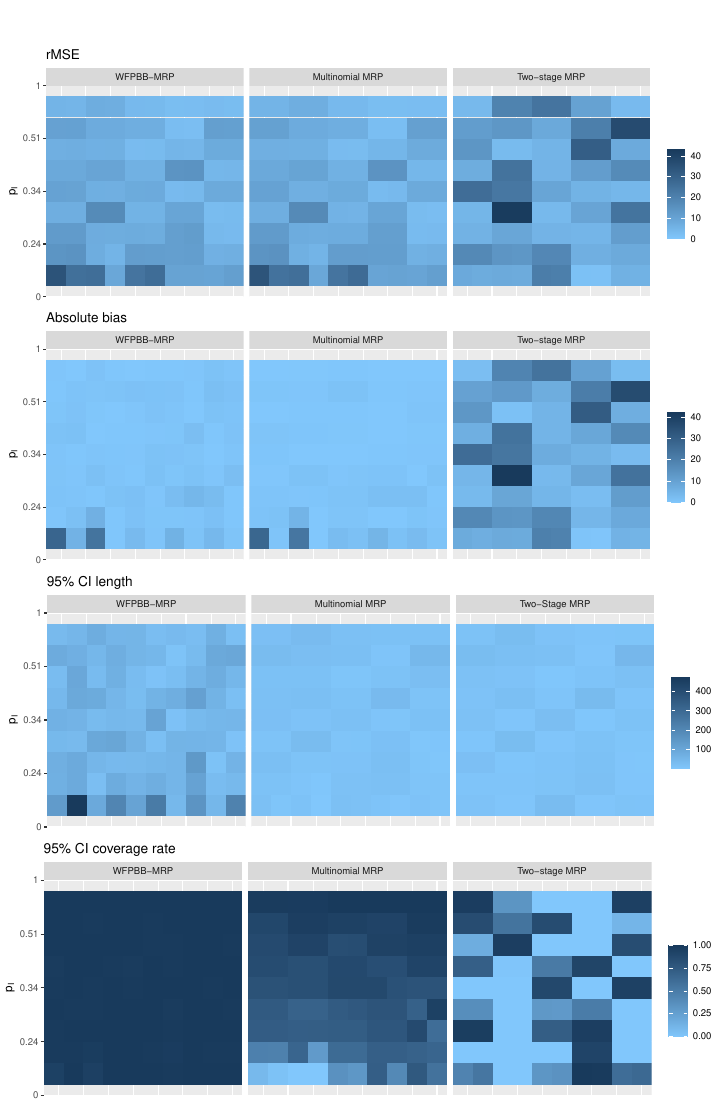}
\caption{Simulation results for the population cell frequency ($\hat{N}_j$) estimates from the EMRP methods (WFPBB-MRP, Multinomial MRP, Two-stage MRP). We report the root mean squared error (rMSE), absolute bias, average 95\% confidence interval (CI) length, and 95\% CI coverage rate. Tiles correspond to cells created by the $(Z,X)$ cross-tabulation and are ordered by inclusion probabilities ($p_I$): cells with smaller inclusion probabilities are at the bottom of the y-axis, with the lowest in the bottom left; cells with greater inclusion probabilities are at the top, with the highest on the upper right. Darker colors correspond to higher values.}
\label{fig:Njint}
\end{figure}

Figure \ref{fig:Njint} presents the performance metrics for the EMRP population cell frequency ($N_j$) estimation in comparison of different methods. Estimating $N_j$ with the WFPBB results in the widest CIs among the EMRP methods for $\hat{N}_j$ and, subsequently, the subgroup estimates. For Two-stage MRP, misspecification of the $(X\mid Z)$ logistic model results in $N_j$ estimates that are substantially more biased than the non-regression methods and contaminates the performance of Two-stage MRP, as shown in Figure~\ref{fig:simresults_subsample}. Multinomial MRP has lower bias values and interval lengths comparable to the Two-stage MRP. However, both methods underestimate the uncertainty in estimating $\hat{N}_j$ and thus lead to low coverage rates of the corresponding EMRP estimators.

Overall, EMRP estimators have higher precision than the design-based methods and smaller bias values than the classical MRP estimator when the distribution $(X\mid Z)$ in the observed sample is different from the population. WFPBB-MRP has larger variances and conservative CI coverage compared to Multinomial MRP and Two-stage MRP.

\section{Application to the Longitudinal Survey of Wellbeing}
\label{application}

The LSW dataset is comprised of two different samples: a phone sample of 2002 residents contacted by random digit dialing, and a face-to-face sample of 226 residents visiting food acquisition agencies for a total of 2228 respondents. We analyze the publicly released data, which do not distinguish the two samples even though they have different sample inclusion mechanisms. Integrating the phone and face-to-face samples will be discussed as a future extension in Section~\ref{discussion}. The study oversamples residents from low-income neighborhoods. The publicly released survey weights have been calibrated to the ACS-NYC 2011 weighted totals and account for unequal probabilities of selection, undercoverage, and nonresponse. While these weights are available in this particular composite sample, the combination is ad hoc and requires improvement with rigorous data integration methods. The frequency of residents visiting a food acquisition agency is related to their food insecurity status; however, we do not have access to the population distribution of the agency visit frequency and will need to estimate it for our analysis.

We classify a respondent to be food insecure if it is ``often''  the case that they ``worried whether [their] food would run out before [they] got money to buy more.'' We use the binary proxy of agency visit frequency to indicate whether they have visited a food acquisition agency within the last 12 months. The outcome model in the EMRP methods accounts for age in years (18-35, 36-50, 50+), sex (male, female), race (White, Black, other), the highest level of education achieved (less than high school, high school or equivalent, some college or associate's degree, bachelor's degree or higher), annual pre-tax cash income for the household ($<$\$35k, \$35-55k, \$55-100k, $>$\$100k), and agency visitor status (``visitor'' if the respondent has visited a food acquisition agency within the last 12 months, ``non-visitor'' otherwise). There are 11 participants with a missing response variable and 20 missing their education values which we impute by randomly sampling from the corresponding observed values, and the effect of the small amount of item nonresponse is negligible.
 
\begin{figure}[ht!]
    \centering
 \begin{tabular}{c}   
\includegraphics[height=3in]{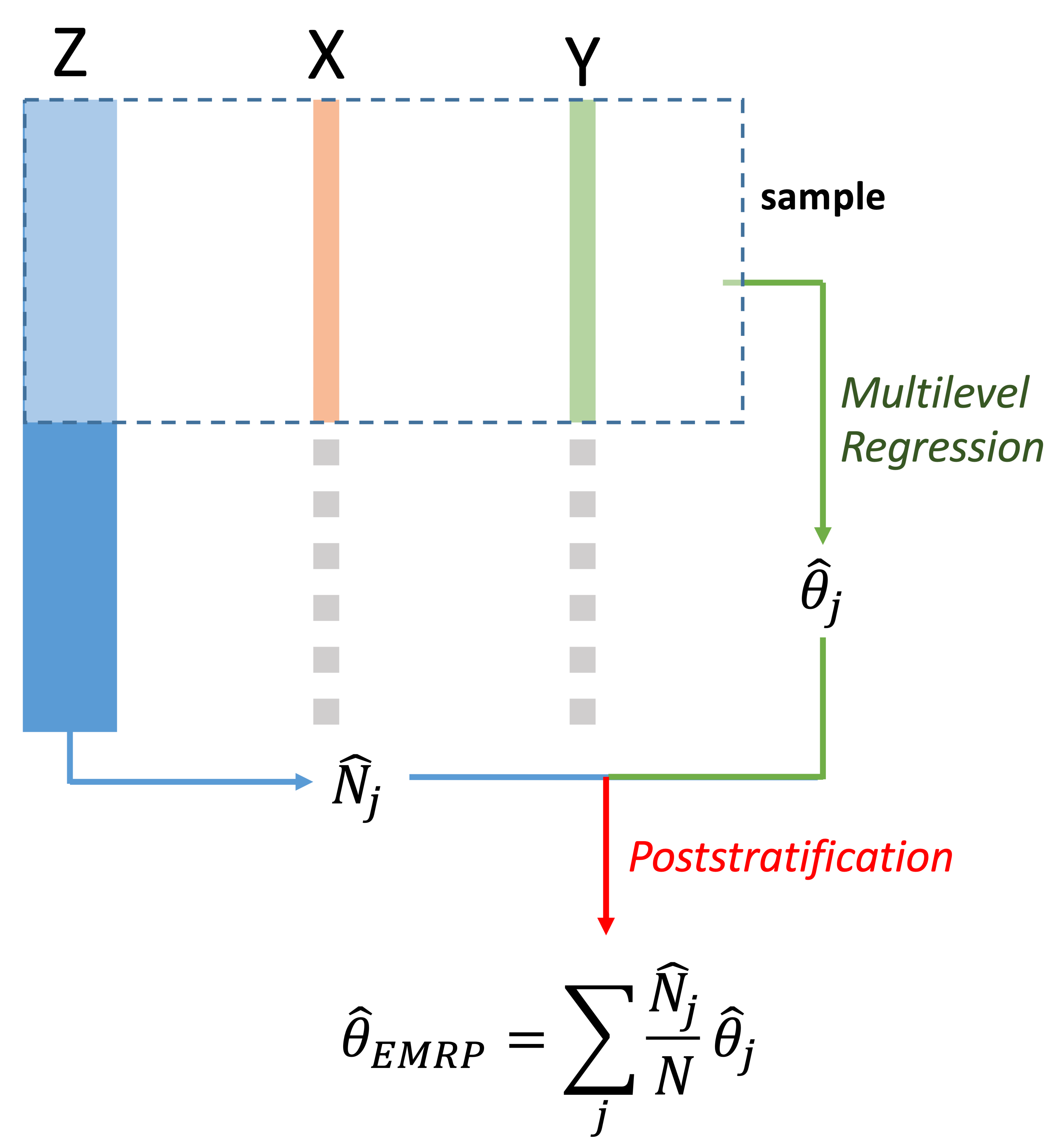}
 \end{tabular}
\caption{Conceptual illustration of the embedded multilevel regression and poststratification (EMRP) workflow.}
 \label{fig:emrp}
  \end{figure}

Table~\ref{tab:baseline_tab1} gives the distributions of sociodemographics and the food insecurity prevalence of the LSW sample and two groups stratified by the indicator of agency visits. Agency visitors tend to be younger, non-White, less educated, and more food insecure compared to non-visitors. Interestingly, the frequency of agency visitors who are low-income residents ($<$\$35k) is lower than that of nonvisitors (32.6\% vs. 57.5\%). The population distribution of sociodemographics (age, sex, race, education, and income) is available in the ACS-NYC 2011, but that of agency visitors is not, reflecting the EMRP setting in Figure~\ref{fig:emrp}.

We apply the EMRP methods to the LSW study to estimate the prevalence of food insecurity among NYC adult residents with different income levels. Two sets of analyses are conducted: one set focuses on the income groups ($<$\$35k, n=884; \$35-55k, n=273; \$55-100k, n=456; and $>$\$100k, n=615) to compare EMRP and classical MRP, and the other set is defined by cross-tabulations of income levels and agency visitor status. In both sets of subgroups, we also include results from directly imputing the outcome with the WFPBB and from the survey-weighted estimator (using publicly released weights). 

For the WFPBB-MRP, we use observed sociodemographic frequencies from the sample ($n^z_m$) and their weighted totals from the ACS ($N^z_m$) to obtain the initial weights in the P\'{o}lya urn $w_{m}=\{N^z_m/n^z_m\}$ and apply WFPBB to estimate $\hat{N}_j$ (where cell $j$ is from the contingency table based on $(Z,X)$) with $L=5000$ synthetic populations and $F=20$ draws from the weighted P\'{o}lya posterior. To reduce the computational burden, we modify the size of each draw $f (=1, \dots, F)$ as $T*n = 30*2228$---which is large enough to overwhelm the sample size---instead of synthesizing the entire population. We set $L$ as a large value of 5000 to match the number of 5000 posterior samples from EMRP. The same parameter settings and base weights are used when directly imputing with the WFPBB.

The outcome model in EMRP includes sociodemographic variables and visit status: 
\[\textrm{logit(Pr}(y_i=1)) =\alpha_0+ \alpha_i^{age} + \alpha_i^{sex} + \alpha_i^{race} + \alpha_i^{educ} + \alpha_i^{income} + \alpha_i^{visit}+\alpha_i^{visit:income},\] 
where $y_i=1$ indicates food insecurity and $y_i=0$ otherwise, for $i=1,\dots,n$. The model includes all main effects and the two-way interaction between visit frequency and income. Classical MRP omits the covariate of visit frequency: $\textrm{logit(Pr}(y_i=1)) =\alpha_0 + \alpha_i^{age} + \alpha_i^{sex} + \alpha_i^{race} + \alpha_i^{educ} + \alpha_i^{income}$, and the same set of covariates are used in the estimation model for visit frequency $(X\mid Z)$ in Two-stage MRP.

We assign the same weakly informative prior distributions to the coefficients as those in Section~\ref{sim-setup}. The model runs four MCMC chains with 10,000 iterations each, with 8750 for burn-in and the last 1250 iterations permuted and merged across chains for a total of 5000 iterations for estimation. The procedures for conducting subgroup inferences from the EMRP and WFPBB estimators are the same as those described in the simulations. The diagnostics indicate model convergence.

 We use Bayesian leave-one-out cross-validation and posterior predictive check to evaluate the goodness of fit,\cite{loo17,gelman:meng:stern96} neither of which raises concerns about model performances. We compare the pointwise out-of-sample prediction accuracy between the EMRP
and MRP models. Pareto k estimates are less than 0.7 for both models, implying that all leave-one-out posteriors are similar to the full posterior. The expected log predictive density for the EMRP model is greater than that of the MRP model (18.4 difference); the EMRP model is preferred for prediction. Details are given in Appendix~\ref{LSWdiag}.

The estimated coefficient for the agency visitor indicator is $\hat{\alpha}^{visit}$ = 0.83 (95\% \textrm{CI}: 0.48, 1.18), and estimates for its interaction terms with income at the $<$\$35k, \$35-55k, \$55-100k, $>$\$100k levels are -0.24 (95\% \textrm{CI}: -1.34, 0.88), 0.31(95\% \textrm{CI}: -0.78, 1.67), 0.54 (95\% \textrm{CI}: -0.44, 1.95), and -0.60 (95\% \textrm{CI}: -2.29, 0.45), respectively. The main effect of $X$ is larger than 0, but the interaction effects have large variability. 

\begin{figure}[ht!]
\centering
\includegraphics[width=0.9\linewidth]{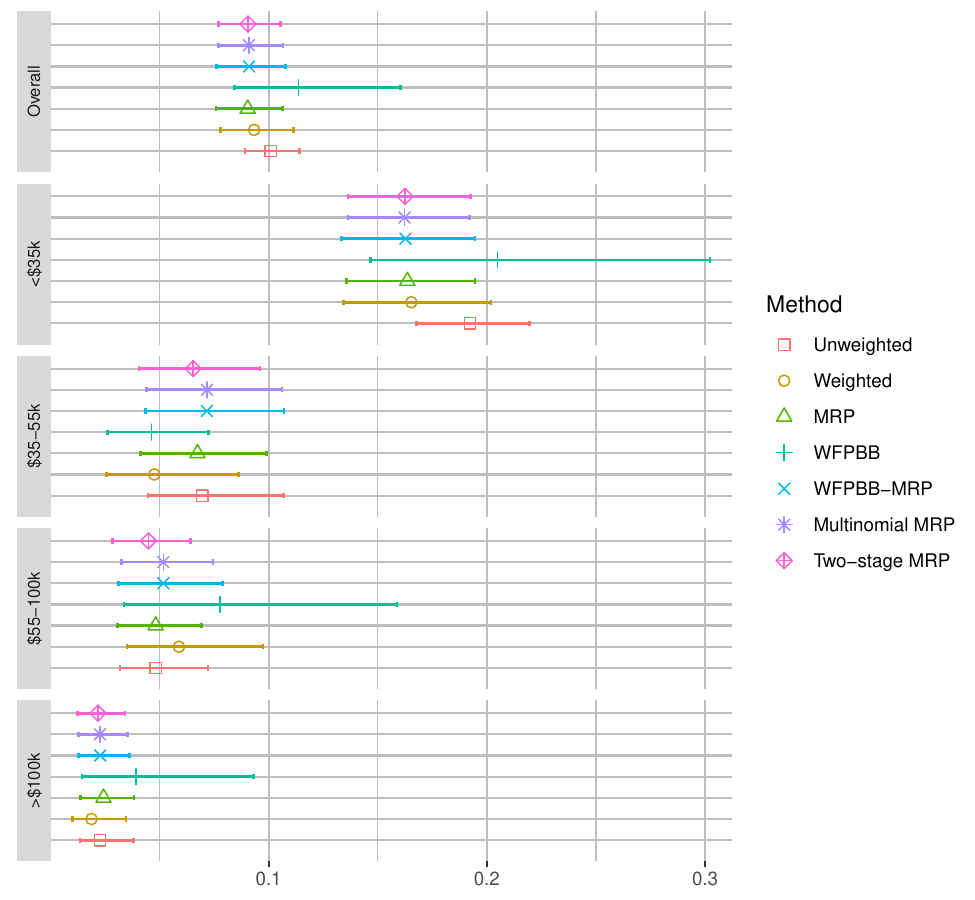}
\caption{Estimates of food insecurity prevalence for the overall NYC adult population and subgroups defined by annual income: $<$\$35k (n=884), \$35-55k (n=273), \$55-100k (n=456), and $>$\$100k (n=615). We report results from the unweighted and survey-weighted estimates, direct imputation with the weighted finite population Bayesian bootstrap (WFPBB), classical multilevel regression and poststratification (MRP), and embedded multilevel regression and poststratification (EMRP) techniques (WFPBB-MRP, Multinomial MRP, Two-stage MRP). Error bars refer to the 95\% confidence intervals.}

\label{fig:forestplot_lswapplied}
\end{figure}

\begin{figure}[ht!]
\centering
\includegraphics[width=0.9\linewidth]{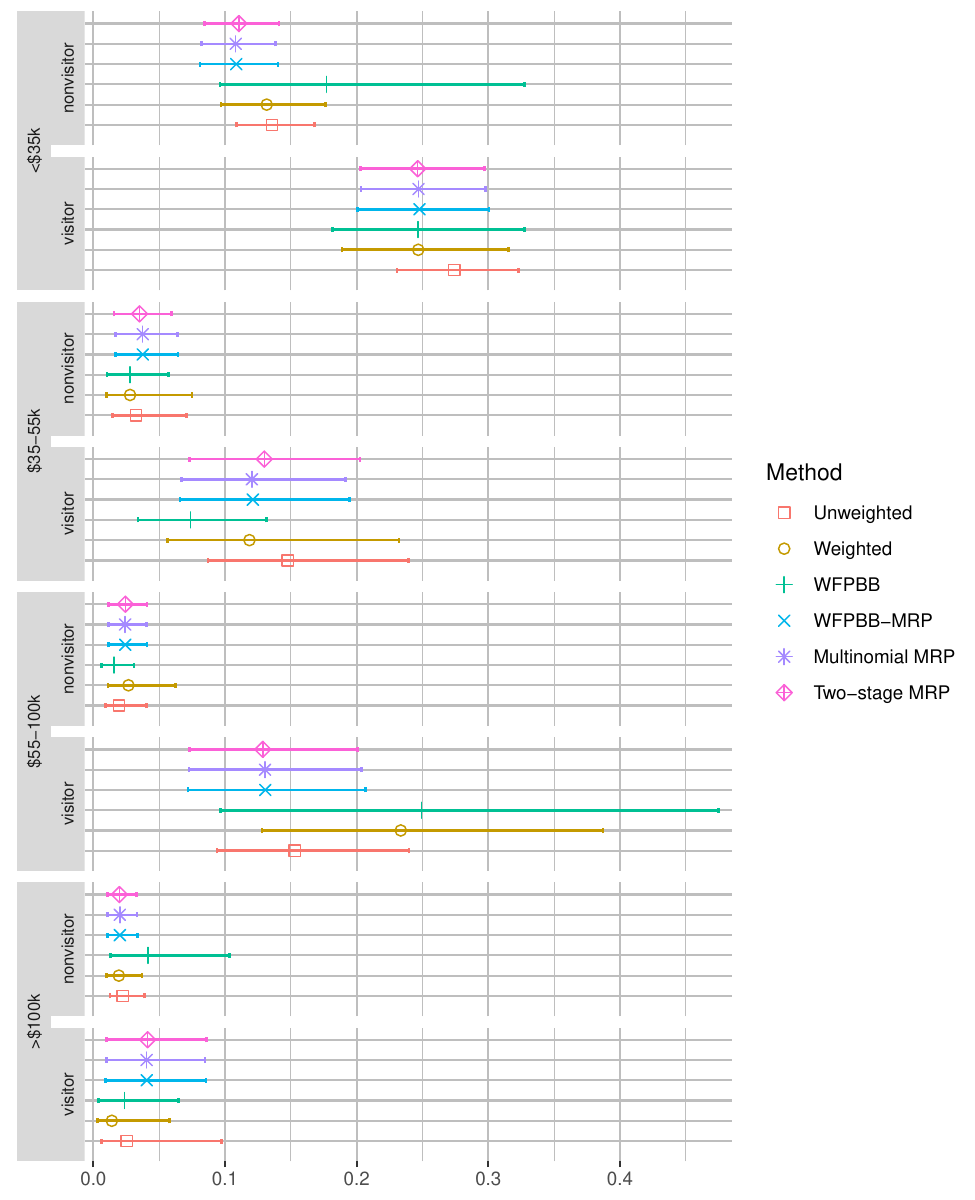}
\caption{Estimates of food insecurity prevalence for the subgroups defined by interactions of annual income and agency visitor status. We report results from the unweighted and survey-weighted estimates, direct imputation with the weighted finite population Bayesian bootstrap (WFPBB), classical multilevel regression and poststratification (MRP), and embedded multilevel regression and poststratification (EMRP) techniques (WFPBB-MRP, Multinomial MRP, Two-stage MRP). Error bars refer to the 95\% confidence intervals.}
\label{fig:forestplot_lswapplied_visitor}
\end{figure}

Figures~\ref{fig:forestplot_lswapplied} and \ref{fig:forestplot_lswapplied_visitor} compare the food insecurity prevalence estimates from the unweighted and survey-weighted analysis, classical MRP, direct imputation of the outcome using the WFPBB, and EMRP for the overall NYC adult population and the four subdomains defined by: 1) annual income intervals and 2) the cross-tabulation between income and agency visit status, respectively. 

In Figure~\ref{fig:forestplot_lswapplied}, the survey-weighted, classical MRP, and EMRP overall prevalence estimates are around 9\%, which is slightly lower than the unweighted estimate of 10.1\% and the WFPBB direct imputation estimate of 11.4\% (see Table~\ref{tab:lswresults} for detailed values). The unweighted estimates are generally higher than the weighted estimates with the exception of the \$55-100k group. The model-based estimates show an inverse relationship between food insecurity and income: around 16\% of those with $<$\$35k annual income are food insecure compared to 2\% of those in the $>$\$100k group. Incorporating weights in the analysis is crucial for bias adjustment. The variances of EMRP estimators are smaller than those of the weighted and WFPBB estimators, where predictive models improve precision. Different EMRP estimators yield similar estimates and mostly overlapping interval widths. Classical MRP point estimates are similar to the EMRP point estimates for the income categories. This is possibly due to small differences in the agent visit frequency between the sample and the population within each income group, the sample sizes of which are large enough.

Figure \ref{fig:forestplot_lswapplied_visitor} presents the prevalence estimates for the subgroups defined by interactions of annual income and agency visitor status. Agency visitors have higher food insecurity compared to those who don't visit, and this gap lessens as annual income increases. The largest gap occurs for individuals with annual income lower than \$35k: approximately 11\% vs. 25\%. Here, the conditional estimates given $(Z, X)$ are substantially different from those only given $Z$. EMRP is beneficial if we are interested in conducting inference on subgroups defined by the missing poststratifier $X$. Agency visitors tend to be more food insecure than the overall study population; the weighted estimate is 20.1\%, more than double the 9.3\% estimated for the overall population (Table~\ref{tab:lswresults-incvisitor}).

The application study shows that accounting for design features is important for correcting the bias in the food insecurity prevalence estimation. The agency visit status is an important predictor of the food insecurity outcome, the distributional imbalance of which between the sample and population within subgroups will affect the mean estimates.

\section{Discussion}
\label{discussion}

Motivated by health disparity research, we focus on estimates for minority groups. MRP has become a popular subgroup estimation method due to its ability to stabilize estimates and adjust for selection bias, but these properties are restricted by whether the population joint distribution of poststratifying auxiliary variables is known. Analysts rarely have access to the population joint distribution of the comprehensive set of predictive auxiliary variables. In such situations, classical use of MRP may require the omission of predictive variables without complete information and produce inaccurate estimates as a result. We have developed the EMRP framework to incorporate variables with incomplete information into MRP by generating synthetic populations to estimate their joint distribution before proceeding with MRP estimation.

Through simulation studies, we compared design-based estimators and direct imputation with the WFPBB with several EMRP methods (WFPBB-MRP, Multinomial MRP, and Two-stage MRP). We found that all EMRP estimators can correct for the bias in classical MRP while maintaining lower standard errors and narrower confidence intervals than directly imputing with the WFPBB and design-based estimates. Performances from the EMRP estimators do not differ substantially from each other, though we would generally recommend the WFPBB-MRP for its consistently high coverage rates. As a benefit of fitting multilevel models and stabilizing small group estimates, the WFPBB-MRP yields bias and rMSE values that are comparable to the Multinomial MRP and Two-stage MRP while producing narrower confidence intervals than directly imputing with the WFPBB. Estimating population cell frequencies using the multinomial distribution leads to small bias values and rMSE for the $\hat{N}_j$ estimates, but low coverage rates of the frequencies in sparsely sampled cells lead to undercoverage for the Multinomial MRP in a few subgroup inferences. Conversely, using MRP to recover $N_j$ as in the Two-stage MRP gives reasonable coverage in most cells, but generates the largest bias and rMSE values among all methods. Model misspecification for $(X\mid  Z)$ can introduce bias in domain inferences, especially for domains with few observations. The WFPBB-MRP avoids this issue by weighting observed cases, accounts for sampling uncertainty when estimating the joint $(Z, X)$ distribution, and combines with MRP to improve the inferences for $(Y \mid  Z, X)$ with a predictive model.

Under settings with incomplete poststratifier $X$ information, differences in bias between classical MRP and EMRP methods are contingent on an imbalance in the missing poststratifying variable's frequency distributions for the inferential subgroup between the sample and population (e.g. a special case is that not all legitimate $X$ values for a given $Z$ in the population are observed in the sample). When the inclusion mechanism is MAR given $Z$, we would expect that the overall mean estimates of MRP and EMRP are similar, but the subgroup estimates could be substantially different. EMRP improves the estimates for subgroups defined by $(Z, X)$. We have compared the differences of the missing poststratifying variable's frequency distributions $\textrm{Pr}(X=1)$ between the population and one randomly drawn sample across the four subgroups constructed under two scenarios in Section~\ref{sim-setup}: 1) based on the joint distribution $(Z, X)$, where four subgroups are based on the percentiles of the inclusion probabilities of the $J$ cells and the distribution of $(X \mid Z)$; and 2) based on only $Z$, where four subgroups by a random sample of 10 cells based on the cross-tabulation of $Z$ variables in each inclusion probability bracket. Table~\ref{x-diff} shows that the first scenario generally has larger differences than the second scenario. The largest difference in the $\textrm{Pr}(X=1)$ is in the group with the lowest inclusion probabilities, for which the EMRP and MRP estimates have the most prominent dissimilarity. If the analyst anticipates that their inferential subgroup will have a balanced $(X\mid Z)$ distribution between the sample and population, then the MRP estimator will have similar bias values to those of the EMRP estimators. The utility of EMRP is best showcased when there is an imbalance of $(X\mid Z)$ in the subgroup between the sample and the population.

The EMRP framework has a few interesting directions for future extensions. First, the EMRP framework can handle general problems of data integration. Datasets from different sources may have incongruous study measures, resulting in incomplete auxiliary information. When combining two datasets obtained through different sampling mechanisms, we modify the $\hat{N}_j$ estimation procedure such that we use only one of the datasets---the one with a selection mechanism that is independent of the incomplete auxiliary variables---to estimate the $(X\mid  Z)$ distribution. To illustrate, the LSW dataset in our application is a composite sample of phone and face-to-face surveys. Given the indicator of the two survey components in the restricted dataset, we would regress $X$ on $Z$ using only the phone sample for the Two-stage MRP, use sociodemographic cell frequencies from the phone sample as the initial sampling weights for the WFPBB-MRP, and use the phone sample to estimate the probabilities for the Multinomial MRP. Integrating data from multiple studies can also present multivariate incomplete auxiliary variables. Such a task would require an iterative or sequential estimation process to estimate the joint distribution, where the choice of WFPBB, multinomial, or MRP models can be tailored to each auxiliary variable and combined under a framework that is similar to multiple imputation. 

Second, in situations where marginal distributions of the multivariate incomplete auxiliary variables are available but their joint distribution is unknown---similar to the raking setting---the WFPBB needs to account for the known constraints when constructing the base weights. Model-based estimation approaches under known margins can be applied.\cite{BayesRake18} 

Third, EMRP can be extended to settings where the data are MNAR because the inclusion mechanism depends on the incomplete auxiliary variables. For example, the inclusion of face-to-face samples in the LSW study depends on the agency visit frequency. When evaluating the performance of EMRP methods under MNAR in our simulation studies, we found that all methods yield bias, but EMRP reduces the bias of classical MRP. Enhancing EMRP methods to handle informative inclusion would further broaden the circumstances under which MRP can be applied successfully.

Finally, the wide use of the EMRP approaches calls for scalable and efficient software development. User-friendly implementations will facilitate broad applications.

\section*{Acknowledgments}
This work is supported by grants from the National Science Foundation (SES1760133) and the National Institutes of Health (U01MD017867). The authors are grateful to Dr. Michael R. Elliott for his assistance in the theory and implementation of the weighted finite population Bayesian bootstrap.

\section*{Data availability statement}

The simulation code is publicly available on GitHub: \url{https://github.com/likat/EMRP}. The data used in the application study are openly available from the New York City Longitudinal Survey of Wellbeing: \url{https://cprc.columbia.edu/content/new-york-city-longitudinal-survey-wellbeing}. 

\subsection*{Financial disclosure}

None reported.

\subsection*{Conflict of interest}

The authors declare no potential conflict of interest.

\section*{Supporting information}
Additional supporting information may be found online in the Appendix section.
\noindent

\bibliography{EMRP}%

\clearpage

\begin{table}[ht!]
\caption{Value ranges of the simulated inclusion probabilities ($p_I$) across cells based on the cross-tabulation of fully observed auxiliary variables.}
\centering
\label{inc-pr}
\begin{tabular}{llllll}
\toprule
\textbf{Cell} & 1--5 & 6--20 & 21--40 & 41--45 & 46--50\\
\textbf{Range of $p_I$} & (0.01, 0.10) & (0.11, 0.40) & (0.21, 0.60) &(0.51, 0.80)&(0.80, 0.99)\\
\bottomrule
\end{tabular}
\end{table}

\begin{table}
\centering
\caption{Differences of the observed values in one random sample and the population values of $\textrm{Pr}(X=1)$ within the subgroups of two simulation cases: 1) subgroup membership is defined based on the joint $(Z,X)$ distribution and 2) membership is defined based on categories of $(Z)$ only. }
\label{x-diff}
	\begin{tabular}{ccccc}
\toprule	
\multirow{2}{*}{$(Z,X)$}	&	$p_I$: 0.03-0.27	&	$p_I$: 0.21-0.38	&	$p_I$: 0.30-0.54	&	$p_I$: 0.40-0.95	\\
\cline{2-5}
	&	0.205	&	-0.046	&	-0.013	&	0.048	\\
	&		&		&		&		\\
\multirow{2}{*}{$(Z)$}	&	$p_I$: 0.03-0.25	&	$p_I$: 0.25-0.38	&	$p_I$: 0.34-0.51	&	$p_I$: 0.40-0.93	\\
\cline{2-5}
	&	0.122	&	-0.031	&	0.003	&	0.054	\\
\bottomrule
\end{tabular}
\end{table}

 \begin{table}[ht!]
  \caption{Simulation results for the overall and subdomain mean estimates, where subdomains are defined by inclusion probability ranges and the joint distribution of $(Z,X)$. We report root mean squared error (rMSE), absolute bias, average 95\% confidence interval (CI) length, and 95\% CI coverage rate from the direct imputation of the outcome using the weighted finite population Bayesian bootstrap (WFPBB), classical multilevel regression and poststratification (MRP), embedded multilevel regression and poststratification (EMRP) methods (WFPBB-MRP, Multinomial MRP, Two-stage MRP), and the unweighted and survey-weighted estimators. }
  \centering
     \begin{tabular}{llcccccc}
 &&Overall & $p_I$: 0.40-0.95 	& $p_I$: 0.30-0.54 &  $p_I$: 0.21-0.38 & $p_I$: 0.03-0.27 \\
	\toprule
Population $\bar{Y}$ &&  0.577 	&  0.581		&	0.427		&	0.584		& 0.618			\\
\midrule
rMSE	&	Unweighted Est. 	&	0.015	&	0.036	&	0.022	&	0.026	&	0.035	\\
	&	Weighted Est. 	&	0.010	&	0.007	&	0.010	&	0.017	&	0.053	\\
	&	WFPBB	&	0.010	&	0.007	&	0.011	&	0.017	&	0.054	\\
	&	Classical MRP	&	0.007	&	0.033	&	0.059	&	0.053	&	0.068	\\
	&	WFPBB-MRP	&	0.007	&	0.007	&	0.011	&	0.011	&	0.018	\\
	&	Multinomial MRP	&	0.007	&	0.007	&	0.012	&	0.011	&	0.018	\\
	&	Two-stage MRP	&	0.007	&	0.008	&	0.013	&	0.014	&	0.017	\\
\midrule													
Bias	&	Unweighted Est. 	&	0.014	&	0.035	&	-0.019	&	-0.019	&	0.016	\\
	&	Weighted Est. 	&	-0.001	&	$<$0.001	&	$<$0.001	&	$<$0.001	&	-0.007	\\
	&	WFPBB	&	$<$0.001	&	-0.001	&	0.001	&	-0.001	&	0.003	\\
	&	Classical MRP	&	-0.002	&	-0.033	&	0.058	&	-0.052	&	-0.067	\\
	&	WFPBB-MRP	&	0.001	&	0.004	&	-0.008	&	-0.005	&	0.007	\\
	&	Multinomial MRP	&	0.000	&	0.004	&	-0.008	&	-0.006	&	0.006	\\
	&	Two-stage MRP	&	-0.002	&	0.006	&	-0.010	&	-0.011	&	0.005	\\
\midrule													
95\% CI length	&	Unweighted Est. 	&	0.023	&	0.035	&	0.046	&	0.068	&	0.104	\\
	&	Weighted Est. 	&	0.035	&	0.038	&	0.047	&	0.069	&	0.164	\\
	&	WFPBB	&	0.045	&	0.050	&	0.061	&	0.088	&	0.202	\\
	&	Classical MRP	&	0.033	&	0.033	&	0.037	&	0.040	&	0.062	\\
	&	WFPBB-MRP	&	0.038	&	0.041	&	0.049	&	0.059	&	0.105	\\
	&	Multinomial MRP	&	0.033	&	0.035	&	0.040	&	0.041	&	0.073	\\
	&	Two-stage MRP	&	0.033	&	0.035	&	0.041	&	0.042	&	0.073	\\
\midrule													
Coverage rate	&	Unweighted Est. 	&	0.305	&	0.005	&	0.640	&	0.805	&	0.835	\\
	&	Weighted Est. 	&	0.945	&	0.990	&	0.975	&	0.955	&	0.880	\\
	&	WFPBB	&	0.950	&	0.995	&	1.000	&	0.990	&	0.940	\\
	&	Classical MRP	&	0.980	&	0.005	&	0.000	&	0.000	&	0.005	\\
	&	WFPBB-MRP	&	1.000	&	1.000	&	0.990	&	0.995	&	1.000	\\
	&	Multinomial MRP	&	0.970	&	0.985	&	0.885	&	0.910	&	0.955	\\
	&	Two-stage MRP	&	0.975	&	0.975	&	0.895	&	0.870	&	0.985	\\
	\bottomrule
	 \end{tabular}
     \label{tab:intxz}
 \end{table}

  \begin{table}[ht!]
  \caption{Simulation results for the overall and subdomain mean estimates, where subdomains are defined by inclusion probability ranges and the levels of $(Z)$. We report root mean squared error (rMSE), absolute bias, average 95\% confidence interval (CI) length, and 95\% CI coverage rate from the direct imputation of the outcome using the weighted finite population Bayesian bootstrap (WFPBB), classical multilevel regression and poststratification (MRP), embedded multilevel regression and poststratification (EMRP) methods (WFPBB-MRP, Multinomial MRP, Two-stage MRP), and the unweighted and survey-weighted estimators. }
  \centering
     \begin{tabular}{llcccccc}
 	&		&	Overall	&	$p_I$: 0.40-0.93	&	$p_I$: 0.34-0.51	&	$p_I$: 0.25-0.38	&	$p_I$: 0.03-0.25	\\
\toprule													
Population $\bar{Y}$	&		&	0.572	&	0.507	&	0.535	&	0.564	&	0.660	\\
\midrule													
rMSE	&	Unweighted Est.	&	0.015	&	0.034	&	0.014	&	0.025	&	0.067	\\
	&	Weighted Est.	&	0.010	&	0.008	&	0.014	&	0.017	&	0.043	\\
	&	WFPBB	&	0.010	&	0.008	&	0.012	&	0.016	&	0.046	\\
	&	Classical MRP	&	0.007	&	0.012	&	0.009	&	0.009	&	0.017	\\
	&	WFPBB-MRP	&	0.007	&	0.007	&	0.010	&	0.011	&	0.019	\\
	&	Multinomial MRP	&	0.007	&	0.007	&	0.010	&	0.011	&	0.019	\\
	&	Two-stage MRP	&	0.007	&	0.011	&	0.009	&	0.011	&	0.017	\\
\midrule													
Bias	&	Unweighted Est.	&	0.014	&	0.033	&	$<$0.001	&	-0.018	&	0.063	\\
	&	Weighted Est.	&	$<$0.001	&	$<$0.001	&	0.002	&	0.001	&	-0.001	\\
	&	WFPBB	&	$<$0.001	&	$<$0.001	&	-0.001	&	$<$0.001	&	0.004	\\
	&	Classical MRP	&	-0.002	&	0.010	&	$<$0.001	&	0.001	&	-0.006	\\
	&	WFPBB-MRP	&	0.001	&	0.005	&	0.001	&	-0.004	&	0.009	\\
	&	Multinomial MRP	&	$<$0.001	&	0.004	&	$<$0.001	&	-0.005	&	0.008	\\
	&	Two-stage MRP	&	-0.002	&	0.009	&	-0.001	&	-0.006	&	$<$0.001	\\
\midrule													
95\% CI length	&	Unweighted Est.	&	0.023	&	0.042	&	0.053	&	0.063	&	0.083	\\
	&	Weighted Est.	&	0.035	&	0.045	&	0.053	&	0.064	&	0.140	\\
	&	WFPBB	&	0.045	&	0.058	&	0.069	&	0.082	&	0.174	\\
	&	Classical MRP	&	0.033	&	0.036	&	0.046	&	0.045	&	0.076	\\
	&	WFPBB-MRP	&	0.038	&	0.046	&	0.051	&	0.057	&	0.104	\\
	&	Multinomial MRP	&	0.033	&	0.036	&	0.046	&	0.045	&	0.077	\\
	&	Two-stage MRP	&	0.033	&	0.036	&	0.046	&	0.046	&	0.078	\\
\midrule													
Coverage rate	&	Unweighted Est.	&	0.340	&	0.055	&	0.935	&	0.775	&	0.170	\\
	&	Weighted Est.	&	0.895	&	1.000	&	0.950	&	0.950	&	0.890	\\
	&	WFPBB	&	0.950	&	0.995	&	0.995	&	1.000	&	0.945	\\
	&	Classical MRP	&	0.980	&	0.935	&	0.990	&	0.980	&	0.975	\\
	&	WFPBB-MRP	&	1.000	&	1.000	&	0.995	&	0.995	&	0.990	\\
	&	Multinomial MRP	&	0.975	&	0.995	&	0.980	&	0.970	&	0.955	\\
	&	Two-stage MRP	&	0.975	&	0.960	&	0.990	&	0.975	&	0.980	\\
\bottomrule													
	 \end{tabular}
     \label{tab:intxz-nosub}
 \end{table}

\begin{table}[ht!]
     \centering
          \caption{Descriptive summary of food insecurity and sociodemographics for respondents from the Longitudinal Survey of Wellbeing, stratified by whether the respondent has visited a food acquisition agency in the last 12 months (visitor) or not (nonvisitor). Values are reported as percentages.}

              \label{tab:baseline_tab1}
     \begin{tabular}{lrrr}
			&	\textbf{Visitor (\%)}	&	\textbf{Nonvisitor (\%)}	&	\textbf{Overall (\%)}	\\
	sample size			&	626		&	1602	&	2228	\\
\toprule
\bf{Age}		&		&		&		\\															
~~~~18-35	&	33.9	&	26.3	&	28.4	\\
~~~~36-50	&	28.0	&	24.7	&	25.6	\\
~~~~50+		&	38.2	&	49.1	&	46.0	\\
\bf{Sex}		&		&		&		\\						
~~~~Male		&	34.8	&	37.1	&	36.4	\\
~~~~Female	&	65.2	&	62.9	&	63.6	\\
\bf{Race} 	&		&		&		\\									
~~~~White	&	13.7	&	37.8	&	31.1	\\
~~~~Black	&	34.8	&	28.3	&	30.1	\\
~~~~Other	&	51.4	&	33.9	&	38.8	\\
\bf{Education}		&		&		&		\\								
~~~~Less than high school				&	22.5	&	11.0	&	14.2	\\
~~~~High school	&	30.4	&	20.1	&	23.0	\\
~~~~Some college				&	27.5	&	22.2	&	23.7	\\
~~~~Bachelors or higher			&	19.6	&	46.7	&	39.1	\\
\bf{Income}		&		&		&		\\											
~~~~$<$\$35k		&	32.6	&	57.7	&	39.7	\\
~~~~\$35-55k		&	11.5	&	14.1	&	12.3	\\
~~~~\$55-100k	&	22.3	&	15.7	&	20.5	\\
~~~~$>$\$100k	&	33.5	&	12.6	&	27.6	\\
\bf{Food security status}	&		&		&		\\											
~~~~Food secure	&	79.4	&	94.0	&	89.9	\\
~~~~Food insecure	&	20.6	&	6.0	&	10.1	\\
\bottomrule
\end{tabular}
 \end{table}

\clearpage

\clearpage
\appendix

\section{Detailed outputs}

\begin{table}[h!]

     \caption{Results from estimating food insecurity prevalence across subgroups defined by annual income brackets: $<$\$35k, n=884; \$35-55k, n=273; \$55-100k, n=456; $>$\$100k, n=615. We compare the point estimate, standard error, and 95\% confidence interval (CI) length between the unweighted and survey-weighted estimators, direct imputation of the outcome using the weighted finite population Bayesian bootstrap (WFPBB), classical multilevel regression and poststratification (MRP), and embedded multilevel regression and poststratification (EMRP) methods (WFPBB-MRP, Multinomial MRP, Two-stage MRP).}
     \centering
     \begin{tabular}{lrrrrr}
 	&	\textbf{Overall}	&	\textbf{$<$\$35k}	&	\textbf{\$35-55k}	&	\textbf{\$55-100k}	&	\textbf{$>$\$100k}	\\
\toprule											
\bf{Estimate}	&		&		&		&		&		\\
~~~~Unweighted Estimate	&	0.101	&	0.192	&	0.070	&	0.048	&	0.023	\\
	&	[0.088, 0.114]	&	[0.167, 0.220]	&	[0.045, 0.107]	&	[0.032, 0.072]	&	[0.014, 0.038]	\\
~~~~Weighted Estimate	&	0.093	&	0.165	&	0.048	&	0.059	&	0.019	\\
	&	[0.078, 0.111]	&	[0.134, 0.202]	&	[0.026, 0.0866]	&	[0.035, 0.097]	&	[0.010, 0.035]	\\
~~~~WFPBB	&	0.114	&	0.205	&	0.046	&	0.078	&	0.039	\\
	&	[0.084, 0.160]	&	[0.147, 0.302]	&	[0.026, 0.073]	&	[0.034, 0.159]	&	[0.014, 0.093]	\\
~~~~Classical MRP	&	0.090	&	0.164	&	0.067	&	0.048	&	0.024	\\
	&	[0.076, 0.106]	&	[0.136, 0.195]	&	[0.041, 0.099]	&	[0.031, 0.069]	&	[0.014, 0.038]	\\
~~~~WFPBB-MRP	&	0.091	&	0.163	&	0.072	&	0.052	&	0.023	\\
	&	[0.076, 0.108]	&	[0.133, 0.195]	&	[0.044, 0.107]	&	[0.031, 0.079]	&	[0.013, 0.036]	\\
~~~~Multinomial MRP	&	0.091	&	0.162	&	0.072	&	0.052	&	0.023	\\
	&	[0.077, 0.107]	&	[0.136, 0.192]	&	[0.044, 0.106]	&	[0.033, 0.074]	&	[0.013, 0.035]	\\
~~~~Two-stage MRP	&	0.091	&	0.162	&	0.065	&	0.045	&	0.022	\\
	&	[0.077, 0.105]	&	[0.136, 0.192]	&	[0.041, 0.096]	&	[0.028, 0.064]	&	[0.012, 0.034]	\\
\midrule											
\bf{Standard error}	&		&		&		&		&		\\
~~~~Unweighted Estimate	&	0.006	&	0.013	&	0.015	&	0.010	&	0.006	\\
~~~~Weighted Estimate	&	0.008	&	0.017	&	0.015	&	0.015	&	0.006	\\
~~~~WFPBB	&	0.020	&	0.040	&	0.012	&	0.033	&	0.021	\\
~~~~Classical MRP	&	0.008	&	0.015	&	0.015	&	0.010	&	0.006	\\
~~~~WFPBB-MRP	&	0.008	&	0.016	&	0.017	&	0.012	&	0.006	\\
~~~~Multinomial MRP	&	0.008	&	0.014	&	0.016	&	0.011	&	0.006	\\
~~~~Two-stage MRP	&	0.007	&	0.014	&	0.014	&	0.009	&	0.006	\\
\midrule											
\bf{95\% CI length}	&		&		&		&		&		\\
~~~~Unweighted Estimate	&	0.025	&	0.052	&	0.060	&	0.039	&	0.024	\\
~~~~Weighted Estimate	&	0.033	&	0.067	&	0.057	&	0.060	&	0.023	\\
~~~~WFPBB	&	0.076	&	0.156	&	0.046	&	0.125	&	0.079	\\
~~~~Classical MRP	&	0.031	&	0.059	&	0.057	&	0.038	&	0.025	\\
~~~~WFPBB-MRP	&	0.032	&	0.061	&	0.063	&	0.047	&	0.023	\\
~~~~Multinomial MRP	&	0.030	&	0.056	&	0.062	&	0.042	&	0.023	\\
~~~~Two-stage MRP	&	0.028	&	0.056	&	0.055	&	0.036	&	0.022	\\
	\bottomrule
	\end{tabular}
     \label{tab:lswresults}
\end{table}

\begin{table}
     \caption{Results from estimating food insecurity prevalence across subgroups defined by interactions of annual income and agency visitor status. We compare the point estimates between the unweighted and survey-weighted estimators, direct imputation of the outcome using the weighted finite population Bayesian bootstrap (WFPBB), and embedded multilevel regression and poststratification (EMRP) methods (WFPBB-MRP, Multinomial MRP, Two-stage MRP) with the marginalized classical multilevel regression and poststratification (MRP) estimate for reference.}
    \footnotesize 
     \centering
     \begin{tabular}[width=0.9\linewidth]{l|cc|cc|cc|cc|cc}
	&	\multicolumn{2}{c|}{\textbf{Overall}}			&	\multicolumn{2}{c|}{\textbf{$<$\$35k}}			&	\multicolumn{2}{c|}{\textbf{\$35-55k}}			&	\multicolumn{2}{c|}{\textbf{\$55-100k}}			&	\multicolumn{2}{c}{\textbf{$>$\$100k}}			\\
	&	Nonvisitor	&	Visitor	&	Nonvisitor	&	Visitor&	Nonvisitor	&	Visitor&	Nonvisitor	&	Visitor&	Nonvisitor	&	Visitor		\\
\toprule																					
Unweighted Estimate	&	0.060	&	0.206	&	0.136	&	0.274	&	0.032	&	0.148	&	0.020	&	0.153	&	0.022	&	0.025	\\
Weighted Estimate	&	0.065	&	0.201	&	0.132	&	0.247	&	0.028	&	0.119	&	0.027	&	0.234	&	0.019	&	0.014	\\
WFPBB	&	0.081	&	0.189	&	0.177	&	0.246	&	0.028	&	0.074	&	0.016	&	0.249	&	0.042	&	0.024	\\
WFPBB-MRP	&	0.053	&	0.178	&	0.109	&	0.248	&	0.038	&	0.121	&	0.024	&	0.131	&	0.020	&	0.041	\\
Multinomial MRP	&	0.053	&	0.178	&	0.108	&	0.247	&	0.038	&	0.120	&	0.024	&	0.130	&	0.020	&	0.040	\\
Two-stage MRP	&	0.054	&	0.191	&	0.111	&	0.246	&	0.035	&	0.130	&	0.024	&	0.129	&	0.020	&	0.041	\\																		
\midrule
	~~~~Classical MRP	&	\multicolumn{2}{c|}{0.090}	&	\multicolumn{2}{c|}{0.164}	&	\multicolumn{2}{c|}{0.067}	&	\multicolumn{2}{c|}{0.048}	&	\multicolumn{2}{c}{0.024}
\end{tabular}
\label{tab:lswresults-incvisitor}																				
\end{table}
\clearpage

\section{Simulation study with only main effects in the $(X \mid Z)$ model}
\label{appendix:main}

 \begin{table}[h!]
    \caption{Simulation results for the overall and subdomain mean estimates, where the $(X \mid Z)$ relationship contains main effects only. Subdomains are defined by inclusion probability ranges and the joint distribution of $(Z, X)$. We report root mean squared error (rMSE), absolute bias, average 95\% confidence interval (CI) length, and 95\% CI coverage rate from the direct imputation of the outcome using the weighted finite population Bayesian bootstrap (WFPBB), classical multilevel regression and poststratification (MRP), embedded multilevel regression and poststratification (EMRP) methods (WFPBB-MRP, Multinomial MRP, Two-stage MRP), and the unweighted and survey-weighted estimators.}
     \centering
     \begin{tabular}{llcccccc}
 &&Overall & $p_I$: 0.40-0.95 	& $p_I$: 0.30-0.54 &  $p_I$: 0.21-0.38 & $p_I$: 0.03-0.27 \\
	\toprule
Population $\bar{Y}$ &&  0.572			& 0.567			&	0.441		&	0.568		& 0.600			\\
\midrule
rMSE	&	Unweighted Est. 	&	0.015	&	0.034	&	0.025	&	0.027	&	0.041	\\
	&	Weighted Est. 	&	0.010	&	0.007	&	0.012	&	0.017	&	0.055	\\
	&	WFPBB	&	0.010	&	0.006	&	0.011	&	0.017	&	0.054	\\
	&	Classical MRP	&	0.007	&	0.048	&	0.076	&	0.060	&	0.081	\\
	&	WFPBB-MRP	&	0.007	&	0.005	&	0.010	&	0.010	&	0.018	\\
	&	Multinomial MRP	&	0.006	&	0.005	&	0.011	&	0.010	&	0.017	\\
	&	Two-stage MRP	&	0.007	&	0.005	&	0.010	&	0.008	&	0.016	\\
\midrule													
Bias	&	Unweighted Est. 	&	0.014	&	0.034	&	-0.022	&	-0.019	&	0.027	\\
	&	Weighted Est. 	&	-0.001	&	$<$0.001	&	$<$0.001	&	0.001	&	-0.006	\\
	&	WFPBB	&	-0.001	&	-0.001	&	0.002	&	-0.003	&	0.004	\\
	&	Classical MRP	&	-0.002	&	-0.048	&	0.075	&	-0.060	&	-0.080	\\
	&	WFPBB-MRP	&	$<$0.001	&	0.002	&	-0.006	&	-0.003	&	0.003	\\
	&	Multinomial MRP	&	$<$0.001	&	0.002	&	-0.007	&	-0.003	&	0.001	\\
	&	Two-stage MRP	&	-0.001	&	0.003	&	-0.007	&	-0.002	&	0.003	\\
\midrule													
95\% CI length	&	Unweighted Est. 	&	0.023	&	0.035	&	0.046	&	0.067	&	0.103	\\
	&	Weighted Est. 	&	0.035	&	0.037	&	0.048	&	0.068	&	0.166	\\
	&	WFPBB	&	0.045	&	0.049	&	0.062	&	0.087	&	0.204	\\
	&	Classical MRP	&	0.033	&	0.033	&	0.037	&	0.040	&	0.062	\\
	&	WFPBB-MRP	&	0.038	&	0.041	&	0.050	&	0.059	&	0.107	\\
	&	Multinomial MRP	&	0.032	&	0.034	&	0.041	&	0.040	&	0.072	\\
	&	Two-stage MRP	&	0.033	&	0.034	&	0.041	&	0.041	&	0.073	\\
\midrule													
Coverage rate	&	Unweighted Est. 	&	0.325	&	0.015	&	0.535	&	0.755	&	0.775	\\
	&	Weighted Est. 	&	0.925	&	0.995	&	0.950	&	0.960	&	0.860	\\
	&	WFPBB	&	0.955	&	1.000	&	1.000	&	0.985	&	0.925	\\
	&	Classical MRP	&	0.985	&	0.000	&	0.000	&	0.000	&	0.000	\\
	&	WFPBB-MRP	&	0.995	&	1.000	&	0.995	&	1.000	&	1.000	\\
	&	Multinomial MRP	&	0.980	&	0.995	&	0.945	&	0.950	&	0.960	\\
	&	Two-stage MRP	&	0.980	&	0.995	&	0.945	&	0.990	&	0.970	\\			\bottomrule
	\end{tabular}
  
     \label{tab:mainxz}
 \end{table}

We consider the case that the $(X\mid Z)$ model under Two-stage MRP is correctly specified. During the simulation study, we assume that Model~\eqref{x-z} only has main effects of $Z$ (MAIN) with
$\beta_0 = -0.5$, $\beta^{Z_a} = (1.7, 0.25, 0.2, -0.75, -1.7)^\top$,  $\beta^{Z_b} = (2.3, 1.5, 0.15, 0.2, 0.9)^\top$, 
$ \beta^{Z_c} = (0, -1)^\top$, and null values for the interaction terms $\beta^{Z_a,Z_c}$ and $\beta^{Z_b,Z_c}$. Other settings are the same as those in Section~\ref{sim-setup}. The simulated population cell frequencies range from 5 to 470, with an average of 100. The average sampled cell sizes in the $(Z, X)$ cross-tabulation table for subgroups $p_I$: 0.03-0.27, $p_I$: 0.21-0.38, $p_I$: 0.30-0.54, and $p_I$: 0.40-0.95 are (9, 24, 50, 89) with total subgroup sample sizes (194, 499, 1017, 1785), respectively. The results are shown in Table~\ref{tab:mainxz}. The conclusions comparing EMRP, MRP and weighted estimators stay the same. Among the three EMRP methods, the CIs from Multinomial MRP and Two-stage MRP are similar and narrower than those under WFPBB-MRP. Multinomial MRP and Two-stage MRP have coverage rates of 94.5\% or above, as an improvement over the case when the $(X\mid Z)$ model is misspecified. The performance of WFPBB-MRP is competitive when Multinomial MRP and Two-stage MRP work well.

\section{Model diagnostics for the application study}
\label{LSWdiag}

\begin{figure}
    \centering
\includegraphics[width=0.9\linewidth]{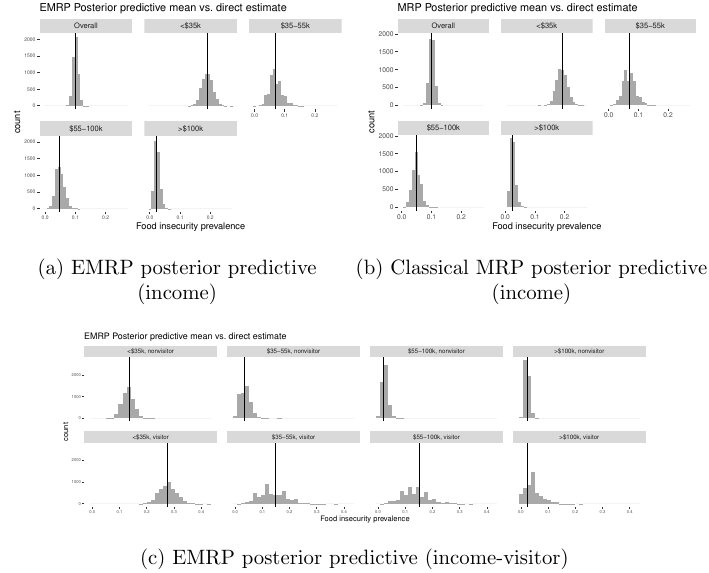}
 \caption{Model diagnostics from applied analysis (posterior predictive distribution and ROC). Posterior predictive distributions are plotted for the food insecurity prevalence for the overall population and (income) or (visitor) analysis sets. Each density consists of 5000 estimates of $\hat{\mu}_k$ based on draws of $\hat{y}$ from the posterior predictive distribution of the LSW outcome model used by either the EMRP or classical MRP methods. The vertical line indicates the value of the direct estimate obtained from the sample.}
 \label{fig:apdiag}
  \end{figure}
\clearpage
We conduct a posterior predictive check and Bayesian leave-one-out cross-validation to evaluate model fit and performance. For the posterior predictive check, we generate posterior predictive distributions of food insecurity prevalence for each analysis subgroup and compare them against the observed values from the LSW sample. Figure \ref{fig:apdiag} includes the histogram of 5000 estimates of $\mu^{pred}_k, pred=1,\ldots,5000$, which are derived from taking the respective subgroup means of the last 5000 posterior predictive outcome draws:

\begin{equation}\label{eq:postpred}
\hat{\mu}^{pred}_k = \frac{1}{n_k}\sum_{i \in k}\hat{y}^{pred}_i.
\end{equation}

The EMRP posterior predictive distributions are centered around the observed values in the relevant subgroups, indicating that the predictive distribution captures the structure of the real data.  

For Bayesian leave-one-out cross-validation, we compare the pointwise out-of-sample prediction accuracy between the EMRP and MRP models. Pareto $k$ estimates are less than 0.7 for both models, which means that all leave-one-out posteriors are similar to the full posterior. The expected log predictive density for the EMRP model is greater than that of the MRP model (18.4 difference); the EMRP model is preferred for prediction.

\end{document}